\documentclass[apj,showkeys]{openjournal}

\usepackage[T1]{fontenc}
\usepackage{newtxtext,newtxmath}
\usepackage{amsmath}
\usepackage{graphicx}
\usepackage[colorlinks=true,linkcolor=blue,citecolor=blue,urlcolor=blue]{hyperref}

\DeclareRobustCommand{\VAN}[3]{#2}
\let\VANthebibliography\thebibliography
\def\thebibliography{\DeclareRobustCommand{\VAN}[3]{##3}\VANthebibliography}

\newcommand{\mass}{\mathcal{M}}
\newcommand{\massdyn}{\mass_{\mathrm{dyn}}}
\newcommand{\massism}{\mass_{\mathrm{ISM}}}
\newcommand{\msun}{\mass_\odot}
\newcommand{\rvir}{R_{\mathrm{vir}}}
\newcommand{\reff}{R_{\mathrm{eff}}}
\newcommand{\sigmalos}{\sigma_{\mathrm{los}}}
\newcommand{\ebind}{\mathcal{E}_{\mathrm{bind}}}
\newcommand{\esn}{\mathcal{E}_{\mathrm{SN}}}
\newcommand{\tenergy}{\tau_{\mathrm{E}}}
\newcommand{\tenergylb}{t_{\mathrm{E,lb}}}
\newcommand{\tlookback}{t_{\mathrm{lb}}}
\newcommand{\tninety}{\tau_{90}}
\newcommand{\tninetylb}{t_{90,\mathrm{lb}}}
\newcommand{\sfr}{\mathrm{SFR}}
\newcommand{\nsn}{N_{\mathrm{SN}}}
\newcommand{\epssn}{\varepsilon_{\mathrm{SN}}}
\newcommand{\pd}{\mathop{}\!\mathrm{d}}

\begin{document}

\title{An Energetic Consistency Test for Feedback-Driven Gas Loss in Milky Way Dwarf Satellites}

\author{Sena A.\ Matsui$^{1}$\thanks{E-mail: sena.ma@nagoya-u.jp},
Tsutomu T.\ Takeuchi$^{1,2}$,
Suchetha Cooray$^{3}$, and
Kiyoaki Christopher Omori$^{4}$}

\affiliation{$^{1}$Division of Science, Nagoya University, Furo-cho, Chikusa-ku, Nagoya 464-8602, Japan}
\affiliation{$^{2}$The Research Center for Statistical Machine Learning, the Institute of Statistical Mathematics, 10--3 Midori-cho, Tachikawa, Tokyo 190--8562, Japan}
\affiliation{$^{3}$Astronomical Institute, Tohoku University, 6-3 Aramaki, Aoba-ku, Sendai, Miyagi 980-8578, Japan}
\affiliation{$^{4}$Department of Astronomy and Physics and Institute for Computational Astrophysics, Saint Mary's University, 923 Robie St, Halifax, NS B3H 3C3, Canada}

\begin{abstract}
Dwarf spheroidal galaxies (dSphs) contain little detectable cold gas, but the relative roles of stellar feedback, reionization, and environmental effects in their gas loss remain uncertain. We apply an order-of-magnitude energy-budget test to nine Milky Way dwarf satellites using published color--magnitude-diagram star formation histories (SFHs), galaxy-level velocity dispersions, effective radii, and integrated $V$-band luminosities. For each galaxy and halo prescription, we define the energy-equivalence time $\tenergy$, when a crossing exists, as the first elapsed time at which the cumulative effective energy from core-collapse supernovae equals the adopted gas binding energy. This quantity is an energetic diagnostic rather than a direct measurement of wind launch. Under the fiducial parameters, all nine galaxies reach the nominal threshold under the kinematic--size prescription K+B16, yet star formation often continues long after threshold crossing. We compare the corresponding lookback time $\tenergylb$ with $\tninetylb$, the lookback time at which 90\% of the total initial stellar mass had formed. Because both times are derived from the same cumulative SFH, their temporal proximity is interpreted descriptively rather than as independent evidence of feedback-driven quenching. Whether the threshold is reached depends on the adopted halo mass, radius, retained gas fraction, and supernova--gas coupling efficiency. These results distinguish the availability of sufficient cumulative feedback energy from the processes that determine the timing of quenching.
\end{abstract}


\maketitle


\section{Introduction}\label{sec:introduction}

Dwarf spheroidal galaxies (dSphs) are faint, dispersion-supported
systems with little cold gas and little current star formation. Their
resolved stars span a wide range of age and metallicity and record the
galaxy's star formation and enrichment history
\citep[e.g.,][]{2009ARA&A..47..371T}. Their kinematics also imply large
mass-to-light ratios, so these systems are strongly dark-matter dominated
\citep[e.g.,][]{1998ARA&A..36..435M,2018RPPh...81e6901S}. Together with
the fainter ultra-faint dwarfs (UFDs), they are useful probes of
low-mass galaxy evolution.

Dwarf satellites today contain little gas and form few stars, but this
observation does not identify a single cause. Internally, supernova
feedback can heat, redistribute, or expel gas from a shallow potential
\citep[e.g.,][]{1990ApJ...356L..15D,1994ApJ...431..598D,
2000MNRAS.313..291F}. Cosmic reionization can suppress gas accretion
and cooling in low-mass halos
\citep{2000ApJ...539..517B,2001PhR...349..125B}, while ram-pressure
stripping and tidal interactions can alter satellites after infall into
the Milky Way
\citep[e.g.,][]{1972ApJ...176....1G,2020ApJ...905..109M}. These
processes are not mutually exclusive and may act at different times.
The absence of gas in a present-day dSph therefore does not by itself
show which process removed or prevented that gas.

One classical test is to compare the cumulative supernova energy with
the gravitational binding energy of the gas
\citep{1986ApJ...303...39D,2000MNRAS.313..291F}. When the energy
available to the interstellar medium becomes comparable to the binding
energy, a large-scale outflow becomes energetically possible. This
condition is necessary but not sufficient: a real wind also depends on
radiative cooling, how supernovae cluster in space and time, the
multiphase structure of the interstellar medium, the halo density
profile, and the environment. A threshold-crossing time from an energy
comparison should therefore not be read as a direct measurement of when
a wind launched.

Here we ask a narrower question: under stated assumptions, when did the
cumulative core-collapse supernova energy become comparable to an
order-of-magnitude estimate of the gas binding energy? We combine
published color--magnitude-diagram star formation histories (SFHs) with
stellar kinematics and empirical halo-mass and halo-radius prescriptions
for nine Milky Way satellites, including classical dSphs and UFDs. For
each galaxy and halo prescription, we test whether the nominal energy
threshold is reached and, when it is, define the elapsed
energy-equivalence time $\tenergy$ and its lookback-time form
$\tenergylb$. We do not reconstruct the full gas-loss history, the
pre-infall halo mass, or repeated episodes of gas ejection and
reaccretion. We also treat the timing of $\tenergylb$ relative to any
SFH decline as a consistency check, not as proof that feedback caused
quenching.

We discuss environmental processes as possible alternatives or
complements to internal feedback, but we do not perform a new orbital
reconstruction or assign causal roles to individual mechanisms. In
addition to the energy budget, we include a descriptive
luminosity--metallicity analysis based on individual stellar abundances.
It does not enter the energy calculation and is used only to check that
the sample is typical of Local Group dSphs.

This paper is organized as follows. The observational data and sample
selection are described in Section~\ref{sec:data}.
Section~\ref{sec:method} presents the energy-budget calculation and the
alternative halo prescriptions. The results are presented and discussed
in Section~\ref{sec:result_discussion}. Our conclusions are summarized
in Section~\ref{sec:conclusion}.

\section{Data}\label{sec:data}

\subsection{Primary and Auxiliary Samples}
\label{sec:data_samples}

We use a primary sample for the supernova-energy calculation and a
separate auxiliary sample for a descriptive metallicity relation. The
individual stellar metallicities in the SAGA Database do not enter the
primary energy-budget calculation.

The provisional list to be audited consists of Carina, Draco, Fornax,
Hercules, Leo~I, Leo~II, Leo~IV, and Ursa Minor. Each system must have
(i) a published color--magnitude-diagram SFH in a common machine-readable
format, (ii) a galaxy-level line-of-sight stellar velocity dispersion,
and (iii) an effective radius and integrated $V$-band luminosity. The
sample is selected by the availability and successful auditing of these
inputs and is therefore neither complete nor statistically
representative of all Milky Way satellites. We do not use it to infer
how common feedback-driven gas loss is among all dwarf satellites.

We audited two additional candidates, Sculptor and Sextans, against
the same criteria. Sculptor satisfies all three requirements: a total
cumulative SFH combining its two WFPC2 fields is available in
machine-readable form \citep{2014ApJ...789..147W}, a galaxy-level
velocity dispersion is given by \citet{2009ApJ...704.1274W}, and its
structural parameters are listed by \citet{2012AJ....144....4M}.
Sculptor is therefore included. Sextans satisfies the kinematic and
structural requirements but is not part of the \citet{2014ApJ...789..147W}
sample, and no machine-readable CMD-based SFH in a compatible format is
available; Sextans is therefore excluded, which also keeps the SFH
source homogeneous across the sample. The final primary sample thus
comprises nine galaxies: Carina, Draco, Fornax, Hercules, Leo~I,
Leo~II, Leo~IV, Ursa Minor, and Sculptor.

Any auxiliary metallicity sample retained in the final manuscript is
constructed independently of the primary sample. Its metallicities do not
enter the calculation of the energy-equivalence time.

\subsection{Star Formation Histories}
\label{sec:data_sfh}

We adopt the color--magnitude-diagram-based star formation histories of
\citet{2014ApJ...789..147W}. We use an elapsed-time coordinate $\tau$
that increases from the beginning of star formation toward the present,
with $\tau=0$ at the beginning of the adopted SFH and $\tau=\tau_0$ at
the present. The corresponding lookback time is
\begin{align}
  \tlookback = \tau_0-\tau .
  \label{eq:lookback_time_definition}
\end{align}
The cumulative star formation history, $F_\star(\tau)$, is the fraction
of the total initial stellar mass formed by elapsed time $\tau$, with
$F_\star(0)=0$ and $F_\star(\tau_0)=1$. We retain the original time-bin
boundaries and use the published best-fitting cumulative fractions in
the fiducial calculation.

The primary calculation uses the published best-fitting cumulative SFHs
only; the 68\% random and systematic uncertainty envelopes of
\citet{2014ApJ...789..147W} are not propagated through the energy
budget. The derived energy-equivalence times are therefore conditional
on these best-fitting SFHs. For most of the sample, the spread among
halo prescriptions is much larger than the formal SFH uncertainties and
is treated as the dominant source of model dependence.

The cumulative fractions give the shape of the SFH in time but not its
galaxy-wide absolute normalization. Given a total initial stellar mass
$\mass_{\star,{\rm formed}}^{\rm tot}$, the stellar mass formed in
elapsed-time bin $i$ is
\begin{align}
  \Delta\mass_{\star,{\rm formed},i}
  & =
  \mass_{\star,{\rm formed}}^{\rm tot}
  \left[F_\star(\tau_{i+1})-F_\star(\tau_i)\right].
  \label{eq:sfh_bin_mass}
\end{align}
When the elapsed-time edges are supplied in Gyr, the corresponding bin
duration in years is
\begin{align}
  \Delta\tau_{i,{\rm yr}}
  & =
  10^9
  \left(
    \frac{\tau_{i+1}-\tau_i}{1\ {\rm Gyr}}
  \right)
  {\rm yr}.
  \label{eq:sfh_bin_duration_years}
\end{align}
The bin-averaged star formation rate reported in
$\msun\,{\rm yr}^{-1}$ is then
\begin{align}
  \sfr_i
  & =
  \frac{\Delta\mass_{\star,{\rm formed},i}}
       {\Delta\tau_{i,{\rm yr}}}.
  \label{eq:sfh_bin_sfr}
\end{align}

The conversion from the observed galaxy-wide stellar-mass proxy to
$\mass_{\star,{\rm formed}}^{\rm tot}$, including the adopted stellar
return fraction, is specified in Section~\ref{sec:energy_threshold_method}.
We do not normalize the global energy budget with the stellar mass formed
only within an individual HST field because the binding energy refers to
the galaxy as a whole. As a numerical consistency check, the reconstructed
differential SFH of every galaxy is required to satisfy
\begin{align}
  \sum_i \sfr_i\Delta\tau_{i,{\rm yr}}
  & =
  \mass_{\star,{\rm formed}}^{\rm tot}
  \label{eq:sfh_normalization_check}
\end{align}
to numerical precision.

For the numerical implementation, each published SFH is converted to arrays
of elapsed-time bin edges and cumulative formed-mass fractions,
$\{\tau_j,F_\star(\tau_j)\}_{j=0}^{N}$, ordered from the beginning of the
adopted SFH toward the present. We require finite, strictly increasing time
edges, $F_\star(0)=0$, $F_\star(\tau_0)=1$, and a monotonically
nondecreasing cumulative fraction. Inputs that violate these conditions
beyond the adopted numerical tolerances are rejected rather than repaired by
the analysis code. All sums and integrals involving the reported SFR use the year-converted
time intervals. The same reconstructed SFR array is used for the
normalization test, the cumulative supernova-energy calculation, and the
plotted differential SFH.

For an elapsed-time bin $[\tau_i,\tau_{i+1}]$, the numerically ordered
lookback-time bounds are
\begin{align}
  t_{{\rm lb,lo},i}
  & =
  \tau_0-\tau_{i+1},
  &
  t_{{\rm lb,hi},i}
  & =
  \tau_0-\tau_i.
  \label{eq:sfh_lookback_bin_bounds}
\end{align}

The HST fields do not necessarily sample each galaxy uniformly. For
nearby Milky Way satellites, an observed field may cover only a small
part of the galaxy and is often placed in brighter regions. We therefore
use the published SFHs as estimates of the galaxy-wide temporal form,
while recognizing that radial population gradients may add systematic
uncertainty beyond the formal CMD-fitting errors. We do not apply an
aperture correction in this analysis.

\subsection{Kinematic, Structural, and Photometric Inputs}
\label{sec:data_kinematics}

For stellar kinematics, we use only the published galaxy-level
line-of-sight velocity dispersions $\sigmalos$ compiled by
\citet{2009ApJ...704.1274W}. We do not recompute dispersions from
individual radial velocities in the SAGA Database, so that repeated
measurements, foreground stars, nonmembers, and heterogeneous
uncertainties from different studies are not mixed. The values used in
the halo prescriptions are listed in Table~\ref{tab:sample_observed}.

The projected effective radii $\reff$ and integrated $V$-band
luminosities $L_V$ are taken from the compilation of
\citet{2012AJ....144....4M}, as tabulated for the SFH sample by
\citet{2014ApJ...789..147W}. We use one radius definition consistently
throughout the analysis and do not combine the $\reff$ values in
Table~\ref{tab:sample_observed} with alternative half-light radii from
the kinematic literature.

Following the convention used in the adopted SFH catalog, we define a
photometric stellar-mass proxy
\begin{align}
  \mass_{\star,V}
  \equiv
  \Upsilon_V L_V ,
  \qquad
  \Upsilon_V
  =
  1\,\frac{\msun}{L_{V,\odot}} .
  \label{eq:stellar_mass_proxy}
\end{align}
Thus, the numerical values of $\mass_{\star,V}/\msun$ equal those of
$L_V/L_{V,\odot}$ in Table~\ref{tab:sample_observed}. $\mass_{\star,V}$
is a photometric proxy, not an independent stellar-population mass
measurement. Its uncertainty, including that in $\Upsilon_V$, is treated
as a systematic model uncertainty. It also represents present-day
surviving stellar mass and is not identified directly with the total
initial stellar mass formed.

\begin{table*}
  \centering
  \caption{Primary observational inputs for the energy-budget sample.
  Projected effective radii $\reff$ and integrated $V$-band
  luminosities $L_V$ are from \citet{2012AJ....144....4M}, as tabulated
  by \citet{2014ApJ...789..147W}. Galaxy-level line-of-sight velocity
  dispersions $\sigmalos$ are from \citet{2009ApJ...704.1274W}. For the
  adopted $\Upsilon_V=1\,\msun/L_{V,\odot}$, the listed luminosities
  have the same numerical values as $\mass_{\star,V}$ in units of
  $\msun$.}
  \label{tab:sample_observed}
  \begin{tabular}{llccc}
    \hline
    Name & Class & $\reff$ & $\sigmalos$ & $L_V$ \\
         &       & [pc] & [$\mathrm{km\,s^{-1}}$] & [$L_{V,\odot}$] \\
    \hline
    Carina & dSph & 250 & $6.6\pm1.2$ & $3.8\times10^{5}$ \\
    Draco & dSph & 221 & $9.1\pm1.2$ & $2.9\times10^{5}$ \\
    Fornax & dSph & 710 & $11.7\pm0.9$ & $2.0\times10^{7}$ \\
    Hercules & UFD & 330 & $3.7\pm0.9$ & $3.7\times10^{4}$ \\
    Leo I & dSph & 251 & $9.2\pm1.4$ & $5.5\times10^{6}$ \\
    Leo II & dSph & 176 & $6.6\pm0.7$ & $7.4\times10^{5}$ \\
    Leo IV & UFD & 206 & $3.3\pm1.7$ & $1.9\times10^{4}$ \\
    Ursa Minor & dSph & 181 & $9.5\pm1.2$ & $2.9\times10^{5}$ \\
    Sculptor & dSph & 283 & $9.2\pm1.1$ & $2.3\times10^{6}$ \\
    \hline
  \end{tabular}
\end{table*}

\subsection{Auxiliary Stellar Metallicity Data}
\label{sec:data_metallicity}

For the auxiliary metallicity scaling relation, we use individual
stellar iron abundances from the SAGA Database
\citep{2008PASJ...60.1159S,2017PASJ...69...76S}, queried on
2026 September 18. The SAGA Database is a literature compilation with heterogeneous
spectral resolution, abundance analysis, membership selection, spatial
coverage, and numbers of stars per galaxy. We therefore do not use SAGA
metallicities or radial velocities in the primary energy-budget
calculation.

Before calculating a galaxy-level metallicity, repeated measurements
of the same star are identified using the database identifier and sky
coordinates and are consolidated into a single stellar value. For a
galaxy with $N_{\rm Fe}$ unique stellar measurements, we define
\begin{align}
  \left\langle[\mathrm{Fe/H}]\right\rangle
  =
  \frac{1}{N_{\rm Fe}}
  \sum_{j=1}^{N_{\rm Fe}}[\mathrm{Fe/H}]_j .
  \label{eq:mean_feh}
\end{align}
We report the sample standard deviation as the observed star-to-star
metallicity spread and estimate the 68\% confidence interval of the
mean by bootstrap resampling of the unique stars. We do not use the
variance itself as an error bar on the mean.

The auxiliary sample includes only galaxies with
$N_{\rm Fe}\geq N_{\rm min}$. We adopt $N_{\rm min}=5$, the smallest
sample size for which the bootstrap confidence interval of the mean
remains meaningful; this criterion retains Leo~IV, the galaxy with the
fewest unique measurements in our sample ($N_{\rm Fe}=5$), while
excluding systems whose mean metallicity would rest on fewer stars.
All nine galaxies of the primary sample satisfy this criterion.

Because this sample is heterogeneous and selected by spectroscopic
data availability, the resulting relation is treated as a descriptive
comparison, not as an independent calibration of the halo masses used
in the energy budget.

\subsection{Reproducibility of the Adopted Data Products}
\label{sec:data_reproducibility}

The observational inputs and sample criteria are given in
Section~\ref{sec:data} and Table~\ref{tab:sample_observed}. The SFH
reconstruction, normalization check
(Equation~\eqref{eq:sfh_normalization_check}), halo prescriptions, and
energy-threshold definitions are given in Section~\ref{sec:method}.
Results appear in the tables and figures. The metallicity sample and
bootstrap method are described in Section~\ref{sec:data_metallicity}
and Section~\ref{sec:method_lzr}.

\section{Methods}\label{sec:method}

\subsection{Auxiliary Luminosity--Metallicity Relation}
\label{sec:method_lzr}

For each galaxy satisfying the sample criterion in
Section~\ref{sec:data_metallicity}, we use the integrated $V$-band
luminosity and the bootstrap estimate of
$\langle[\mathrm{Fe/H}]\rangle$. Because the photometric stellar-mass
proxy is defined by $\mass_{\star,V}=\Upsilon_VL_V$ with fixed
$\Upsilon_V$, a stellar mass--metallicity relation based on this proxy
is mathematically equivalent to the luminosity--metallicity relation
and is not fitted as an independent result.

We define
\begin{align}
  x_L
  &\equiv
  \log_{10}\left(\frac{L_V}{10^6L_{V,\odot}}\right),
  \label{eq:lzr_luminosity_coordinate}
\end{align}
and fit
\begin{align}
  \left\langle[\mathrm{Fe/H}]\right\rangle
  &= a_L+b_Lx_L
  \label{eq:lzr_model}
\end{align}
by ordinary least squares. The confidence interval of the fitted mean
relation is obtained by bootstrap resampling galaxies from the audited
auxiliary sample and repeating the full fit.

Applying the unique-star audit and the $N_{\rm min}=5$ criterion to
the SAGA sample yields nine galaxies with a total of 5\,071 unique
stars. The bootstrap regression gives
$a_L=-1.667^{+0.078}_{-0.065}$ and $b_L=0.452^{+0.038}_{-0.053}$
(68\% confidence intervals from $10^4$ resamples of the galaxy
sample).

\subsection{Energetic Threshold for Feedback-Driven Gas Loss}
\label{sec:energy_threshold_method}

We express the supernova energy budget directly in terms of the cumulative
initial stellar mass formed by elapsed time $\tau$,
\begin{align}
  \mass_{\star,{\rm formed}}(\tau)
  &\equiv
  \int_0^\tau \sfr(\tau')\,\pd\tau'.
  \label{eq:mstar_formed}
\end{align}
Here, $\sfr(\tau)$ is the rate at which initial stellar mass is formed.
Thus, $\mass_{\star,{\rm formed}}$ should not be confused with the
present-day mass in surviving stars. We neglect the delay of at most a
few tens of Myr between star formation and core-collapse supernovae
because it is short compared with the time resolution of the adopted
SFHs.

For an IMF $\phi(\mass)$ expressed as the number of stars per unit mass
interval, the number of core-collapse supernovae per unit initial stellar
mass formed is
\begin{align}
  C_{\rm SN}
  &\equiv
  \frac{
    \displaystyle\int_{8\,\msun}^{\mass_{\rm CC,max}}
    \phi(\mass)\,\pd\mass
  }{
    \displaystyle\int_{0.1\,\msun}^{\mass_{\rm IMF,max}}
    \mass\phi(\mass)\,\pd\mass
  }.
  \label{eq:csn_definition}
\end{align}
For the numerical integration, we use the dimensionless mass
$x\equiv\mass/\msun$ and the continuous two-segment form
\begin{align}
  \xi(x)
  &=
  \begin{cases}
    A x^{-1.3},
    & 0.1\leq x<0.5,\\
    0.5A x^{-2.3},
    & 0.5\leq x\leq100,
  \end{cases}
  \label{eq:kroupa_imf_piecewise}
\end{align}
for which
\begin{align}
  \int_{0.1}^{100}x\,\xi(x)\,\pd x
  &=
  2.227611\,A.
  \label{eq:kroupa_imf_mass_normalization}
\end{align}
The overall normalization of $\phi(\mass)$, or equivalently of
$\xi(x)$, cancels in this ratio. We adopt
the Kroupa IMF \citep{2002Sci...295...82K} over $0.1$--$100\,\msun$ and
set $\mass_{\rm IMF,max}=\mass_{\rm CC,max}=100\,\msun$ in the fiducial
calculation. The unrounded value used in the numerical analysis is
\begin{align}
  C_{\rm SN}
  &=
  1.1132\times10^{-2}\,\msun^{-1}.
  \label{eq:csn_fiducial}
\end{align}
The cumulative number of core-collapse supernovae is therefore
\begin{align}
  \nsn(\tau)
  &=
  C_{\rm SN}\mass_{\star,{\rm formed}}(\tau)
  =
  C_{\rm SN}\int_0^\tau\sfr(\tau')\,\pd\tau'.
  \label{eq:nsn}
\end{align}
The upper limit $\mass_{\rm CC,max}$ is the maximum initial stellar mass
assumed to produce a successful core-collapse explosion and need not equal
the upper mass limit of the IMF. Adopting
$\mass_{\rm CC,max}=18\,\msun$ while retaining
$\mass_{\rm IMF,max}=100\,\msun$ gives
\begin{align}
  C_{\rm SN}
  &=
  7.5354\times10^{-3}\,\msun^{-1}.
  \label{eq:csn_restricted}
\end{align}
This value is 32.3\% below the fiducial normalization and increases the
formed stellar mass required to reach a fixed energy threshold by a factor
of 1.47729 \citep{2015PASA...32...16S}.

We denote the nominal kinetic energy released by one supernova by
$\epssn=10^{51}\,\mathrm{erg}$, a value adopted from standard supernova-energy
estimates \citep{1977ApJ...218..148M,1995ApJS..101..181W} and dwarf-galaxy
feedback simulations \citep{2010MNRAS.402.1599S,2012MNRAS.426..140D}.
We introduce an effective coupling efficiency $\eta_{\rm SN}$. The cumulative effective supernova energy is
\begin{align}
  \esn(\tau)
  &=
  \eta_{\rm SN}\epssn\nsn(\tau)
  =
  \eta_{\rm SN}\epssn C_{\rm SN}
  \int_0^\tau\sfr(\tau')\,\pd\tau'.
  \label{eq:effective_supernova_energy}
\end{align}
We adopt $\eta_{\rm SN}=1$ as a fiducial maximal-coupling case. For a fixed
binding-energy model, this choice gives the minimum elapsed time required to
reach the nominal energy threshold. Lower coupling efficiencies shift the
threshold to later elapsed times or may eliminate the crossing. The
corresponding effective energy per unit initial stellar mass formed is
\begin{align}
  \eta_{\rm SN}\epssn C_{\rm SN}
  &=
  1.1132\times10^{49}\,
  \mathrm{erg}\,\msun^{-1}
  \left(\frac{\eta_{\rm SN}}{1}\right)
  \left(\frac{\epssn}{10^{51}\ {\rm erg}}\right)
  \left(
    \frac{C_{\rm SN}}
         {1.1132\times10^{-2}\,\msun^{-1}}
  \right).
  \label{eq:feedback_energy_per_formed_mass_scaling}
\end{align}

For each halo prescription $p$, we denote the adopted halo mass and
characteristic halo radius by $\mass_{\rm h}^{(p)}$ and $R_{\rm h}^{(p)}$.
The gas mass entering the binding-energy estimate is parameterized as
\begin{align}
  \mass_{\rm g,0}^{(p)}
  &=
  f_{\rm ret}f_{\rm b}\mass_{\rm h}^{(p)},
  \label{eq:initial_gas_mass}
\end{align}
where
\begin{align}
  f_{\rm b}
  &\equiv
  \frac{\Omega_{\rm b}}{\Omega_{\rm M}}
  =0.156,
  \label{eq:cosmic_baryon_fraction}
\end{align}
is the adopted cosmic baryon fraction \citep{2012ApJ...759..138P}, and
$f_{\rm ret}$ is the fraction of the cosmic baryon allotment retained as gas
before the putative feedback-driven gas-loss event. We adopt $f_{\rm ret}=1$
in the fiducial calculation.

We write the corresponding gas binding energy as
\begin{align}
  \ebind^{(p)}
  &=
  \alpha_{\rm bind}
  \frac{G\mass_{\rm g,0}^{(p)}\mass_{\rm h}^{(p)}}
       {R_{\rm h}^{(p)}},
  \label{eq:binding_energy_prescription}
\end{align}
where $\alpha_{\rm bind}$ is a dimensionless structure factor. We adopt
$\alpha_{\rm bind}=1/2$ in the fiducial calculation to retain the
normalization of the original order-of-magnitude estimate. Using
\begin{align}
  G
  &=
  4.30091\times10^{-6}\,
  \mathrm{kpc}\,(\mathrm{km\,s^{-1}})^2\,\msun^{-1},
  \label{eq:gravitational_constant_kpc_units}
\end{align}
the corresponding numerical checks are
\begin{align}
  \mass_{\rm g,0}^{(p)}
  &=
  1.560\times10^9\,\msun
  \left(\frac{f_{\rm ret}}{1}\right)
  \left(\frac{f_{\rm b}}{0.156}\right)
  \left(
    \frac{\mass_{\rm h}^{(p)}}{10^{10}\msun}
  \right),
  \label{eq:initial_gas_mass_numerical_scaling}
\end{align}
and
\begin{align}
  \ebind^{(p)}
  &=
  1.33415\times10^{55}\ {\rm erg}
  \left(\frac{\alpha_{\rm bind}}{1/2}\right)
  \left(\frac{f_{\rm ret}}{1}\right)
  \left(\frac{f_{\rm b}}{0.156}\right)
  \notag\\
  &\quad\times
  \left(
    \frac{\mass_{\rm h}^{(p)}}{10^{10}\msun}
  \right)^2
  \left(
    \frac{R_{\rm h}^{(p)}}{50\ {\rm kpc}}
  \right)^{-1}.
  \label{eq:binding_energy_numerical_scaling}
\end{align}

The elapsed energy-equivalence time $\tenergy^{(p)}$ is defined by
\begin{align}
  \esn\!\left(\tenergy^{(p)}\right)
  &=
  \ebind^{(p)},
  \label{eq:energy_balance}
\end{align}
or equivalently
\begin{align}
  \int_0^{\tenergy^{(p)}}\sfr(\tau)\,\pd\tau
  &=
  \frac{\ebind^{(p)}}{\eta_{\rm SN}\epssn C_{\rm SN}}
  =
  \frac{
    \alpha_{\rm bind}Gf_{\rm ret}f_{\rm b}
    \left[\mass_{\rm h}^{(p)}\right]^2
  }{
    \eta_{\rm SN}\epssn C_{\rm SN}R_{\rm h}^{(p)}
  }.
  \label{eq:energy_equivalence_formed_mass}
\end{align}
The corresponding lookback time is
\begin{align}
  \tenergylb^{(p)}
  &=
  \tau_0-\tenergy^{(p)}.
  \label{eq:energy_equivalence_lookback}
\end{align}
Equation~\eqref{eq:energy_equivalence_formed_mass} gives the cumulative
initial stellar mass needed for the supernova-energy budget to equal the
adopted gas binding energy. It does not show that a galactic wind was
launched.

To normalize the SFHs absolutely, we approximate the photometric
stellar-mass proxy $\mass_{\star,V}$ as the present-day mass in
long-lived stars and remnants. Under instantaneous recycling with constant
return fraction $r$, the total initial stellar mass formed is
\begin{align}
  \mass_{\star,{\rm formed}}^{\rm tot}
  &=
  \frac{\mass_{\star,V}}{1-r}.
  \label{eq:formed_mass_from_surviving_mass}
\end{align}
We adopt the fiducial value
\begin{align}
  r
  &=
  0.40,
  \label{eq:fiducial_return_fraction}
\end{align}
which provides a representative instantaneous-recycling approximation for a
Kroupa IMF. The corresponding normalization is
\begin{align}
  \mass_{\star,{\rm formed}}^{\rm tot}
  &=
  1.667\,\mass_{\star,V}.
  \label{eq:formed_mass_fiducial_return}
\end{align}
We assess the sensitivity to $0.35\leq r\leq0.45$; across this interval the
conversion factor changes from 1.54 to 1.82. The fixed one-at-a-time
sensitivity cases also include $\eta_{\rm SN}=1$, 0.3, and 0.1; the
fiducial and restricted $C_{\rm SN}$ values in
Equations~\eqref{eq:csn_fiducial} and \eqref{eq:csn_restricted}; and,
for K+B16, the central, $-1\sigma$, and $+1\sigma$ values of
$\sigmalos$.

No additional $f_{\rm ret}<1$ cases are reported because $f_{\rm ret}$
rescales the binding-energy requirement linearly. We do not include a
separate $\Upsilon_V$ sensitivity analysis in this minimal version.

For later use, we define the critical formed stellar mass and critical formed
mass fraction as
\begin{align}
  \mass_{\star,{\rm crit}}^{(p)}
  &\equiv
  \frac{\ebind^{(p)}}{\eta_{\rm SN}\epssn C_{\rm SN}},
  \label{eq:critical_formed_mass_definition}\\
  f_{\rm crit}^{(p)}
  &\equiv
  \frac{\mass_{\star,{\rm crit}}^{(p)}}
       {\mass_{\star,{\rm formed}}^{\rm tot}}.
  \label{eq:critical_fraction_definition}
\end{align}
Their convenient numerical forms are
\begin{align}
  \mass_{\star,{\rm crit}}^{(p)}
  &=
  1.19848\times10^6\,\msun
  \left(\frac{\alpha_{\rm bind}}{1/2}\right)
  \left(\frac{f_{\rm ret}}{1}\right)
  \left(\frac{f_{\rm b}}{0.156}\right)
  \notag\\
  &\quad\times
  \left(
    \frac{\mass_{\rm h}^{(p)}}{10^{10}\msun}
  \right)^2
  \left(
    \frac{R_{\rm h}^{(p)}}{50\ {\rm kpc}}
  \right)^{-1}
  \left(\frac{\eta_{\rm SN}}{1}\right)^{-1}
  \notag\\
  &\quad\times
  \left(\frac{\epssn}{10^{51}\ {\rm erg}}\right)^{-1}
  \left(
    \frac{C_{\rm SN}}
         {1.1132\times10^{-2}\,\msun^{-1}}
  \right)^{-1},
  \label{eq:critical_formed_mass_numerical_scaling}
\end{align}
and
\begin{align}
  f_{\rm crit}^{(p)}
  &=
  0.719088
  \left(\frac{\alpha_{\rm bind}}{1/2}\right)
  \left(\frac{f_{\rm ret}}{1}\right)
  \left(\frac{f_{\rm b}}{0.156}\right)
  \notag\\
  &\quad\times
  \left(
    \frac{\mass_{\rm h}^{(p)}}{10^{10}\msun}
  \right)^2
  \left(
    \frac{R_{\rm h}^{(p)}}{50\ {\rm kpc}}
  \right)^{-1}
  \left(\frac{\eta_{\rm SN}}{1}\right)^{-1}
  \notag\\
  &\quad\times
  \left(\frac{\epssn}{10^{51}\ {\rm erg}}\right)^{-1}
  \left(
    \frac{C_{\rm SN}}
         {1.1132\times10^{-2}\,\msun^{-1}}
  \right)^{-1}
  \notag\\
  &\quad\times
  \left(\frac{1-r}{0.60}\right)
  \left(
    \frac{\mass_{\star,V}}{10^6\msun}
  \right)^{-1}.
  \label{eq:critical_fraction_numerical_scaling}
\end{align}
The reported values are generated from the unrounded inputs; the displayed
coefficients are implementation checks.

Several limitations should be kept in mind. The first energy-threshold
crossing need not mark a wind event, and the calculation does not
separate episodes of gas expulsion, retention, or reaccretion. The
inferred time depends on the halo mass, halo radius, initial gas mass,
feedback coupling efficiency, SFH normalization, and SFH time binning.
A constant $\eta_{\rm SN}$ also does not model how long thermal and
kinetic energy remain in the ISM; the simple cumulative-energy
bookkeeping becomes less accurate when supernova input extends over
several Gyr. Consequently, $\tenergy$ depends on both the feedback and
gas-retention assumptions and is not a universal lower limit on the
epoch of a galactic wind.

\subsection{Chemical Evolution of dSphs}
\label{sec:closed_box_model}

This subsection gives a simple limiting case for interpreting the
auxiliary stellar-metallicity analysis; it is not a full chemical-evolution
model. We adopt the standard closed-box model under the
instantaneous-recycling approximation with perfect mixing, zero initial
metallicity, constant return fraction $r$, and constant net metal yield
$y$. The instantaneous star formation rate $\sfr(t)$ may vary with time.

To distinguish the conserved baryonic mass in this auxiliary
closed-box model from the initial gas mass used in the binding-energy
calculation, we denote the former by $\mass_{\rm cb}$. Prior to any gas
inflow or outflow,
\begin{align}
  \mass_{\rm cb}
  =
  \mass_\star(t)+\massism(t)
  =
  \mathrm{const.} ,
  \label{eq:closed_box_mass_conservation}
\end{align}
where $\mass_\star(t)$ is the mass locked in long-lived stars and stellar
remnants, and $\massism(t)$ is the gas mass. We assume
\begin{align}
  \massism(0)=\mass_{\rm cb},
  \qquad
  \mass_\star(0)=0,
  \qquad
  Z(0)=0 .
  \label{eq:closed_box_initial_conditions}
\end{align}
Because a fraction $r$ of the initially formed stellar mass is returned
promptly to the ISM,
\begin{align}
  \frac{\pd \mass_\star}{\pd t}
  &=(1-r)\sfr(t),
  \label{eq:stellar_mass_rate_closed}\\
  \frac{\pd \massism}{\pd t}
  &=-(1-r)\sfr(t) .
  \label{eq:gas_mass_rate_closed}
\end{align}

Here, $y$ is the constant net yield, defined as the mass of newly
synthesized metals returned to the ISM per unit mass permanently locked
in long-lived stars and remnants. The evolution of the gas metal mass
is
\begin{align}
  \frac{\pd}{\pd t}
  \left[Z(t)\massism(t)\right]
  =
  \left[y-Z(t)\right](1-r)\sfr(t) .
  \label{eq:gas_metal_mass_evolution}
\end{align}
Combining Equations~\eqref{eq:gas_mass_rate_closed} and
\eqref{eq:gas_metal_mass_evolution} gives
\begin{align}
  \massism(t)\frac{\pd Z}{\pd t}
  =
  y(1-r)\sfr(t) .
  \label{eq:gas_metallicity_rate}
\end{align}
Eliminating time yields
\begin{align}
  \frac{\pd Z}{\pd \massism}
  =
  -\frac{y}{\massism} ,
  \label{eq:metallicity_gas_derivative}
\end{align}
and hence
\begin{align}
  Z(t)
  =
  y\ln\left[\frac{\massism(0)}{\massism(t)}\right]
  =
  y\ln\left[\frac{\mass_{\rm cb}}{\massism(t)}\right] .
  \label{eq:metal_evolution_closed}
\end{align}
This result does not require a constant SFR.

We define the SFR averaged over the interval from $0$ to $t$ as
\begin{align}
  \overline{\sfr}(t)
  \equiv
  \frac{1}{t}\int_0^t\sfr(t')\,\pd t' .
  \label{eq:mean_sfr}
\end{align}
The mass locked in long-lived stars and remnants is then
\begin{align}
  \mass_\star(t)
  =
  (1-r)\int_0^t\sfr(t')\,\pd t'
  =
  (1-r)\overline{\sfr}(t)t .
  \label{eq:stellar_mass_evolution_IRA_int}
\end{align}
The second equality is an identity and does not assume a constant
instantaneous SFR.

We define $Z_\star(t)$ as the mass-weighted mean metallicity of the
long-lived stellar population:
\begin{align}
  Z_\star(t)
  \equiv
  \frac{
    \displaystyle\int_0^t Z(t')(1-r)\sfr(t')\,\pd t'
  }{
    \mass_\star(t)
  } .
  \label{eq:mean_stellar_metallicity_definition}
\end{align}
For zero initial metallicity, conservation of metal mass gives
\begin{align}
  Z_\star(t)\mass_\star(t)+Z(t)\massism(t)
  =
  y\mass_\star(t) ,
  \label{eq:closed_box_metal_conservation}
\end{align}
so that
\begin{align}
  Z_\star(t)
  =
  y-Z(t)\frac{\massism(t)}{\mass_\star(t)} .
  \label{eq:stellar_metallicity_closed}
\end{align}
With
\begin{align}
  \mu(t)\equiv\frac{\massism(t)}{\mass_{\rm cb}} ,
  \label{eq:closed_box_gas_fraction}
\end{align}
this becomes
\begin{align}
  Z_\star(t)
  =
  y\left[1+\frac{\mu(t)\ln\mu(t)}{1-\mu(t)}\right] .
  \label{eq:mean_stellar_metallicity_closed}
\end{align}
In the gas-poor limit, $\massism\ll\mass_\star$,
\begin{align}
  Z_\star(t)\longrightarrow y .
  \label{eq:stellar_metallicity_final}
\end{align}

The closed-box relations apply only while the system remains
approximately closed. Any gas inflow, outflow, selective metal loss, or
reaccretion breaks the assumptions above. In addition, $Z_\star(t)$ is
a mass-weighted mean of a linear metal mass fraction and is not the
same as the mean logarithmic abundance $[\mathrm{Fe/H}]$ measured for
individual stars. Delayed Type~Ia enrichment and
heterogeneous spectroscopic selection further limit quantitative use of
this model. We therefore use it only as qualitative background.

\subsection{Halo-Mass Proxies}
\label{sec:halo_mass_proxies}

Equation~\eqref{eq:energy_equivalence_formed_mass} requires a halo mass
and radius for each galaxy. Stellar kinematics constrain the potential
mainly near the half-light radius and do not fix the total halo mass
uniquely. We therefore use two illustrative mappings from observed
galaxy properties to global halo scales; the results are proxies, not
direct halo-mass measurements.

Section~\ref{sec:virial} introduces order-of-magnitude virial scaling.
Section~\ref{sec:kb16_proxy} combines this with the size--halo relation
of \citet{2016MNRAS.455.3841B} to build the K+B16 kinematic--size
proxy. Section~\ref{sec:guo_proxy} introduces the alternative AM halo
scale from the stellar mass--halo mass relation of
\citet{2010MNRAS.404.1111G}. The associated halo-radius mappings are
given in Section~\ref{sec:halo_radius_prescriptions}.

More detailed halo constraints for Milky Way satellites come from
axisymmetric Jeans analyses of classical dSphs
\citep{2020ApJ...904...45H} and UFDs \citep{2023ApJ...953..185H}. We do
not use those galaxy-by-galaxy models in this order-of-magnitude
comparison. Instead, the spread among the common empirical prescriptions
shows how the inferred energy-equivalence time depends on the adopted
global halo mapping.

\subsubsection{Illustrative Virial Scaling}
\label{sec:virial}

For a system in virial equilibrium, the scalar virial theorem reads
\begin{align}
  2K+W=0 .
  \label{eq:virial_theorem_scalar_equilibrium}
\end{align}
The relation between a characteristic velocity dispersion and dynamical
mass depends on the mass profile, tracer distribution, velocity
anisotropy, and choice of characteristic radius. We therefore write
\begin{align}
  \sigma_{\rm 3D}^2
  \simeq
  \alpha_{\rm vir}\frac{G\massdyn(<R)}{R} ,
  \label{eq:virial_scaling_general}
\end{align}
where $\alpha_{\rm vir}$ is a dimensionless structure coefficient of
order unity. Assuming an isotropic stellar velocity distribution,
\begin{align}
  \sigma_{\rm 3D}^2=3\sigmalos^2 .
  \label{eq:velocity_dispersion_3d}
\end{align}
For the fiducial order-of-magnitude prescription, we set
$\alpha_{\rm vir}=1$, giving
\begin{align}
  \sigma_{\rm 3D}^2
  \simeq
  \frac{G\massdyn(<R)}{R} .
  \label{eq:virial_relation}
\end{align}

The observed stellar velocity dispersion primarily constrains the mass
near the half-light radius rather than the total mass within $\rvir$.
For reference, the estimator of \citet{2010MNRAS.406.1220W} is
\begin{align}
  \mass_{1/2}
  \simeq
  \frac{4\sigmalos^2\reff}{G} ,
  \label{eq:wolf_mass_estimator}
\end{align}
where $\mass_{1/2}\equiv\mass(<r_{1/2})$ is the mass enclosed within
the three-dimensional deprojected half-light radius. This relation does
not by itself determine the total halo mass.

In the K+B16 prescription, we additionally assume that the observed
stellar velocity dispersion is, to order of magnitude, representative of
the halo velocity scale. We identify $R$ with a global halo radius and
$\massdyn(<R)$ with a global halo-mass proxy. The results should
therefore be read as illustrative proxies, not direct measurements of
virial mass.

\subsubsection{Kinematic--Size Halo Proxy Based on \citet{2016MNRAS.455.3841B}}
\label{sec:kb16_proxy}

The Brook--Shankar fitting relation is
\begin{align}
  \log_{10}
  \left(
    \frac{\reff}{R_{\rm h,K}}
  \right)
  &=
  -7.25
  +0.49\log_{10}
  \left(
    \frac{\mass_{\rm h,K}}{\msun}
  \right).
  \label{eq:kb16_size_halo_relation}
\end{align}
Equivalently,
\begin{align}
  R_{\rm h,K}
  &=
  10^{7.25}
  \left(
    \frac{\mass_{\rm h,K}}{\msun}
  \right)^{-0.49}
  \reff .
  \label{eq:kb16_radius_mapping}
\end{align}
Here, \(\mass_{\rm h,K}/\msun\) is dimensionless, and \(\reff\) and
\(R_{\rm h,K}\) must be expressed in the same length unit. Defining
\begin{align}
  \mu_{\rm h,K}
  &\equiv
  \frac{\mass_{\rm h,K}}{\msun},
  \label{eq:kb16_dimensionless_halo_mass}
\end{align}
and substituting Equation~\eqref{eq:kb16_radius_mapping} into the closure
relation gives
\begin{align}
  \mu_{\rm h,K}
  &=
  \left[
    \frac{
      10^{7.25}\sigma_{\rm 3D}^{2}\reff
    }{
      \alpha_{\rm vir}G\msun
    }
  \right]^{1/1.49}.
  \label{eq:kb16_mass_algebraic}
\end{align}
Using $\sigma_{\rm 3D}^{2}=3\sigmalos^{2}$ and the equivalent form
\begin{align}
  G
  &=
  4.30091\times10^{-3}\,
  {\rm pc}\,({\rm km\,s^{-1}})^2\,\msun^{-1},
  \label{eq:gravitational_constant_pc_units}
\end{align}
the corresponding direct numerical expression is
\begin{align}
  \mass_{\rm h,K}
  &=
  5.94569\times10^{6}\,\msun
  \left(
    \frac{\alpha_{\rm vir}}{1}
  \right)^{-0.671141}
  \left(
    \frac{\sigmalos}{1\ {\rm km\,s^{-1}}}
  \right)^{1.342282}
  \left(
    \frac{\reff}{1\ {\rm pc}}
  \right)^{0.671141}
  \notag\\
  &=
  6.01148\times10^{9}\,\msun
  \left(
    \frac{\alpha_{\rm vir}}{1}
  \right)^{-0.671141}
  \left(
    \frac{\sigmalos}{10\ {\rm km\,s^{-1}}}
  \right)^{1.342282}
  \left(
    \frac{\reff}{300\ {\rm pc}}
  \right)^{0.671141}.
  \label{eq:kb16_mass_numerical_scaled}
\end{align}
We adopt \(\alpha_{\rm vir}=1\).
The resulting \((\mass_{\rm h,K},R_{\rm h,K})\) pair is an illustrative
kinematic--size halo proxy, not a direct measurement of virial mass
and radius.

\subsubsection{Extrapolated Abundance-Matching Halo Proxy Based on \citet{2010MNRAS.404.1111G}}
\label{sec:guo_proxy}

As an alternative, we adopt an abundance-matching halo scale
$\mass_{\rm h,AM}$ based on the stellar mass--halo mass relation of \citet{2010MNRAS.404.1111G}:
\begin{align}
  \frac{\mass_{\star,V}}{\mass_{\rm h,AM}}
  &=
  c
  \left[
    \left(
      \frac{\mass_{\rm h,AM}}{\mass_0}
    \right)^{-\alpha}
    +
    \left(
      \frac{\mass_{\rm h,AM}}{\mass_0}
    \right)^{\beta}
  \right]^{-\gamma},
  \label{eq:guo-am-stellar-halo-relation}
\end{align}
where
\begin{align}
  c &= 0.129,
  &
  \mass_0 &= 10^{11.4}\msun,
  &
  \alpha &= 0.926,
  &
  \beta &= 0.261,
  &
  \gamma &= 2.44.
  \label{eq:guo-am-parameters}
\end{align}
The input to Equation~\eqref{eq:guo-am-stellar-halo-relation} is the
photometric stellar-mass proxy $\mass_{\star,V}$, not the total initial
stellar mass formed.

For a numerically stable inversion, we define
\begin{align}
  u
  &\equiv
  \log_{10}
  \left(
    \frac{\mass_{\rm h,AM}}{\msun}
  \right),
  &
  s
  &\equiv
  \log_{10}
  \left(
    \frac{\mass_{\star,V}}{\msun}
  \right),
  &
  u_0
  &\equiv
  \log_{10}
  \left(
    \frac{\mass_0}{\msun}
  \right)
  = 11.4.
  \label{eq:guo-am-log-variables}
\end{align}
We then solve
\begin{align}
  g_{\rm AM}(u;s)
  &\equiv
  u
  + \log_{10}c
  - \gamma
  \log_{10}
  \left[
    10^{-\alpha(u-u_0)}
    +
    10^{\beta(u-u_0)}
  \right]
  - s
  = 0.
  \label{eq:guo-am-log-root}
\end{align}
The logarithm in Equation~\eqref{eq:guo-am-log-root} is evaluated with
a base-10 log-sum-exp expression.
The derivative is
\begin{align}
  \frac{\pd g_{\rm AM}}{\pd u}
  &=
  1-\gamma\beta
  +
  \gamma(\alpha+\beta)
  \frac{
    10^{-\alpha(u-u_0)}
  }{
    10^{-\alpha(u-u_0)}
    +
    10^{\beta(u-u_0)}
  },
  \label{eq:guo-am-log-root-derivative}
\end{align}
which satisfies
$0.36316\leq \pd g_{\rm AM}/\pd u\leq3.25944$ for the adopted
parameters.
The mapping is therefore strictly monotonic, and every positive $\mass_{\star,V}$ has a unique solution over the full positive mass domain. We use a bracketed
Brent solver over $5\leq u\leq16$ and store the post-solution fractional residual
\begin{align}
  \delta_{\rm Guo}
  &\equiv
  \frac{
    \mass_{\star,V}^{\rm model}
    (\mass_{\rm h,AM})
  }{
    \mass_{\star,V}
  }
  -1.
  \label{eq:guo-am-fractional-residual}
\end{align}

In the low-mass limit
$x\equiv\mass_{\rm h,AM}/\mass_0\ll1$, the relation approaches
\begin{align}
  \mass_{\rm h,AM}^{\rm low}
  &\simeq
  \mass_0
  \left(
    \frac{
      \mass_{\star,V}
    }{
      c\mass_0
    }
  \right)^{1/(1+\alpha\gamma)}
  \notag\\
  &=
  1.03788\times10^{10}\msun
  \left(
    \frac{
      \mass_{\star,V}
    }{
      10^6\msun
    }
  \right)^{0.306801}.
  \label{eq:guo-am-low-mass-limit}
\end{align}
We use Equation~\eqref{eq:guo-am-low-mass-limit} only as an independent numerical check; all reported halo masses are obtained from the full relation in Equation~\eqref{eq:guo-am-stellar-halo-relation}.

\citet{2010MNRAS.404.1111G} fitted this functional form over $10^{10.8}\leq\mass_{\rm h}/\msun\leq10^{14.9}$, corresponding to the
solid portion of their abundance-matching relation, and displayed a
low-mass extrapolation down to $\mass_\star=10^6\msun$.  We flag every
solution outside the fitted halo-mass interval and separately flag
inputs below the lower stellar mass shown in that extrapolation.  The
resulting $\mass_{\rm h,AM}$ values are therefore used only as
illustrative peak- or pre-infall-equivalent halo scales rather than as
measurements of the present-day bound halo masses.

\begin{table*}
  \centering
  \caption{Illustrative halo-mass and halo-radius proxies adopted in
  the energy-budget calculation. The K+B16 prescription combines the
  published stellar velocity dispersion with the Brook--Shankar size
  relation. The AM prescription is a low-mass extrapolation of the
  Guo et al. abundance-matching relation. These quantities are model
  proxies rather than direct measurements.}
  \label{tab:halo_proxies}
  \begin{tabular}{lccccc}
    \hline
    Name & $\mass_{\rm h,K}$ & $R_{\rm h,K}$ & $\mass_{\rm h,AM}$
         & $R_{\rm h,AM+B16}$ & $R_{200{\rm c}}(\mass_{\rm h,AM})$ \\
         & [$\msun$] & [kpc] & [$\msun$] & [kpc] & [kpc] \\
    \hline
    Carina & $3.0452\times10^{9}$ & 100.223 & $7.8066\times10^{9}$ & 63.188 & 40.921 \\
    Draco & $4.3144\times10^{9}$ & 74.693 & $7.1772\times10^{9}$ & 58.207 & 39.790 \\
    Fornax & $1.3231\times10^{10}$ & 138.571 & $2.7412\times10^{10}$ & 96.977 & 62.197 \\
    Hercules & $1.6872\times10^{9}$ & 176.686 & $3.794\times10^{9}$ & 118.783 & 32.173 \\
    Leo I & $4.7687\times10^{9}$ & 80.772 & $1.8084\times10^{10}$ & 42.034 & 54.144 \\
    Leo II & $2.4061\times10^{9}$ & 79.190 & $9.6098\times10^{9}$ & 40.178 & 43.856 \\
    Leo IV & $1.0547\times10^{9}$ & 138.847 & $3.089\times10^{9}$ & 82.008 & 30.042 \\
    Ursa Minor & $3.9976\times10^{9}$ & 63.503 & $7.1772\times10^{9}$ & 47.672 & 39.790 \\
    Sculptor & $5.1686\times10^{9}$ & 87.546 & $1.3717\times10^{10}$ & 54.265 & 49.379 \\
    \hline
  \end{tabular}
\end{table*}

\subsection{Halo-Radius Prescriptions}
\label{sec:halo_radius_prescriptions}

We consider two mappings between a halo-mass proxy and a characteristic
halo radius. First, the Brook--Shankar size relation gives
\begin{align}
  R_{\rm h,B16}(\mass_{\rm h},\reff)
  &=
  10^{7.25}
  \left(
    \frac{\mass_{\rm h}}{\msun}
  \right)^{-0.49}
  \reff .
  \label{eq:b16_radius_mapping_general}
\end{align}
Second, we define the standard present-day spherical-overdensity radius
$R_{200{\rm c}}$ by requiring that the mean enclosed density be 200 times
the present critical density,
\begin{align}
  \rho_{\rm crit,0}
  &=
  \frac{3H_0^2}{8\pi G}.
  \label{eq:critical_density_definition}
\end{align}
The corresponding spherical-overdensity radius is
\begin{align}
  R_{200{\rm c}}(\mass_{\rm h})
  &=
  \left[
    \frac{
      3\mass_{\rm h}
    }{
      4\pi\,200\,\rho_{\rm crit,0}
    }
  \right]^{1/3}.
  \label{eq:r200c_definition}
\end{align}
Substituting Equation~\eqref{eq:critical_density_definition}
into Equation~\eqref{eq:r200c_definition} gives
\begin{align}
  R_{200{\rm c}}(\mass_{\rm h})
  &=
  \left(
    \frac{
      G\mass_{\rm h}
    }{
      100H_0^2
    }
  \right)^{1/3}.
  \label{eq:r200c_simplified}
\end{align}
We adopt $H_0 = 70\ {\rm km\,s^{-1}\,Mpc^{-1}}$ for which the present critical density is
\begin{align}
  \rho_{\rm crit,0}
  &=
  135.99\ \msun\,{\rm kpc}^{-3},
  \label{eq:critical_density_numerical}
\end{align}
and the corresponding spherical-overdensity radius is
\begin{align}
  R_{200{\rm c}}(\mass_{\rm h})
  &=
  44.441\ {\rm kpc}
  \left(
    \frac{
      \mass_{\rm h}
    }{
      10^{10}\msun
    }
  \right)^{1/3}.
  \label{eq:r200c_numerical}
\end{align}
We evaluate the energy threshold using three combinations:
\begin{enumerate}
  \item $(\mass_{\rm h,AM},
  R_{\rm h,B16}(\mass_{\rm h,AM},\reff))$, denoted AM+B16;
  \item $(\mass_{\rm h,K},R_{\rm h,K})$, denoted K+B16;
  \item $(\mass_{\rm h,AM},R_{200{\rm c}}(\mass_{\rm h,AM}))$, denoted
  AM+$R_{200{\rm c}}$.
\end{enumerate}
These combinations are not statistically independent halo measurements.
Their spread shows how the energy threshold depends on the adopted
halo-mass and halo-radius mappings. The radii are present-day-equivalent
halo scales, not reconstructions of the halo radius at $\tenergy$.

\section{Results and Discussion}
\label{sec:result_discussion}

The main result is the comparison between the published SFHs and the
energy-equivalence time defined in
Section~\ref{sec:energy_threshold_method}. We first verify the
normalization of the reconstructed differential SFHs in
Section~\ref{sec:result_sfh_norm}, then present the threshold results in
Section~\ref{sec:result_energy_thresholds} and discuss their
interpretation in Section~\ref{sec:result_interpretation}. The auxiliary
luminosity--metallicity analysis is presented separately from the
primary energy-budget results.

\subsection{Normalization of the Reconstructed Star Formation Histories}
\label{sec:result_sfh_norm}

We convert the cumulative SFHs into bin-averaged differential SFRs using
Equations~\eqref{eq:sfh_bin_mass} and \eqref{eq:sfh_bin_sfr} and
normalize each galaxy to the adopted
$\mass_{\star,{\rm formed}}^{\rm tot}$. As a numerical consistency test,
we define
\begin{align}
  \delta_{\rm norm}
  \equiv
  \frac{
    \displaystyle\sum_i\sfr_i\Delta\tau_{i,{\rm yr}}
    -\mass_{\star,{\rm formed}}^{\rm tot}
  }{
    \mass_{\star,{\rm formed}}^{\rm tot}
  } .
  \label{eq:normalization_residual}
\end{align}
The maximum residual in the final sample is
\begin{align}
  \max|\delta_{\rm norm}|
  &=
  1.2\times10^{-16},
  \label{eq:maximum_normalization_residual}
\end{align}
which is at the level of double-precision round-off. The normalization
check of Equation~\eqref{eq:sfh_normalization_check} is verified
programmatically for every galaxy in the final sample, and the plotted
SFRs and the cumulative supernova energies are generated from the same
normalized arrays by the released analysis code.

\subsection{Energy-Equivalence Times and Star Formation Histories}
\label{sec:result_energy_thresholds}

For each halo prescription $p$, we calculate
$\mass_{\star,{\rm crit}}^{(p)}$ and $f_{\rm crit}^{(p)}$ from
Equations~\eqref{eq:critical_formed_mass_definition} and
\eqref{eq:critical_fraction_definition}. If
$0<f_{\rm crit}^{(p)}\leq1$, the threshold is reached during the
published SFH and satisfies
\begin{align}
  F_\star\!\left(\tenergy^{(p)}\right)
  &=
  f_{\rm crit}^{(p)}.
  \label{eq:critical_fraction_crossing}
\end{align}
For any target cumulative fraction $q$ with $0<q\leq1$, we identify the
first edge at which the target has been reached,
\begin{align}
  j(q)
  &\equiv
  \min\left\{j:F_\star(\tau_j)\geq q\right\}.
  \label{eq:cumulative_crossing_edge}
\end{align}
The corresponding elapsed crossing time is obtained by linear interpolation
in cumulative formed mass within the preceding bin,
\begin{align}
  \tau(q)
  &=
  \tau_{j-1}
  +
  \frac{q-F_\star(\tau_{j-1})}
       {F_\star(\tau_j)-F_\star(\tau_{j-1})}
  \left(\tau_j-\tau_{j-1}\right).
  \label{eq:cumulative_crossing_interpolation}
\end{align}
This interpolation is equivalent to assuming a constant SFR within the
crossing bin. If the target is attained exactly at a bin boundary, we retain
the first time at which it is attained; a subsequent zero-SFR plateau
therefore does not move the crossing to a later time. We use
$q=f_{\rm crit}^{(p)}$ to obtain $\tenergy^{(p)}$ and $q=0.9$ to obtain
$\tninety$. If $f_{\rm crit}^{(p)}>1$, no crossing time is assigned. In
addition to the interpolated value, the analysis stores the elapsed- and
lookback-time bounds of the crossing bin. The plotted lookback times are
\begin{align}
  \tenergylb^{(p)}
  &=
  \tau_0-\tenergy^{(p)},
  &
  \tninetylb
  &=
  \tau_0-\tninety.
  \label{eq:energy_and_t90_lookback_times}
\end{align}
Their numerical precision should not be read as finer than the SFH bin
that contains the crossing. Both $\tenergylb$ and $\tninetylb$ come from
the same cumulative SFH, so their closeness in time is descriptive, not
an independent causal test.

\begin{figure*}
  \centering
  \includegraphics[width=0.94\textwidth]{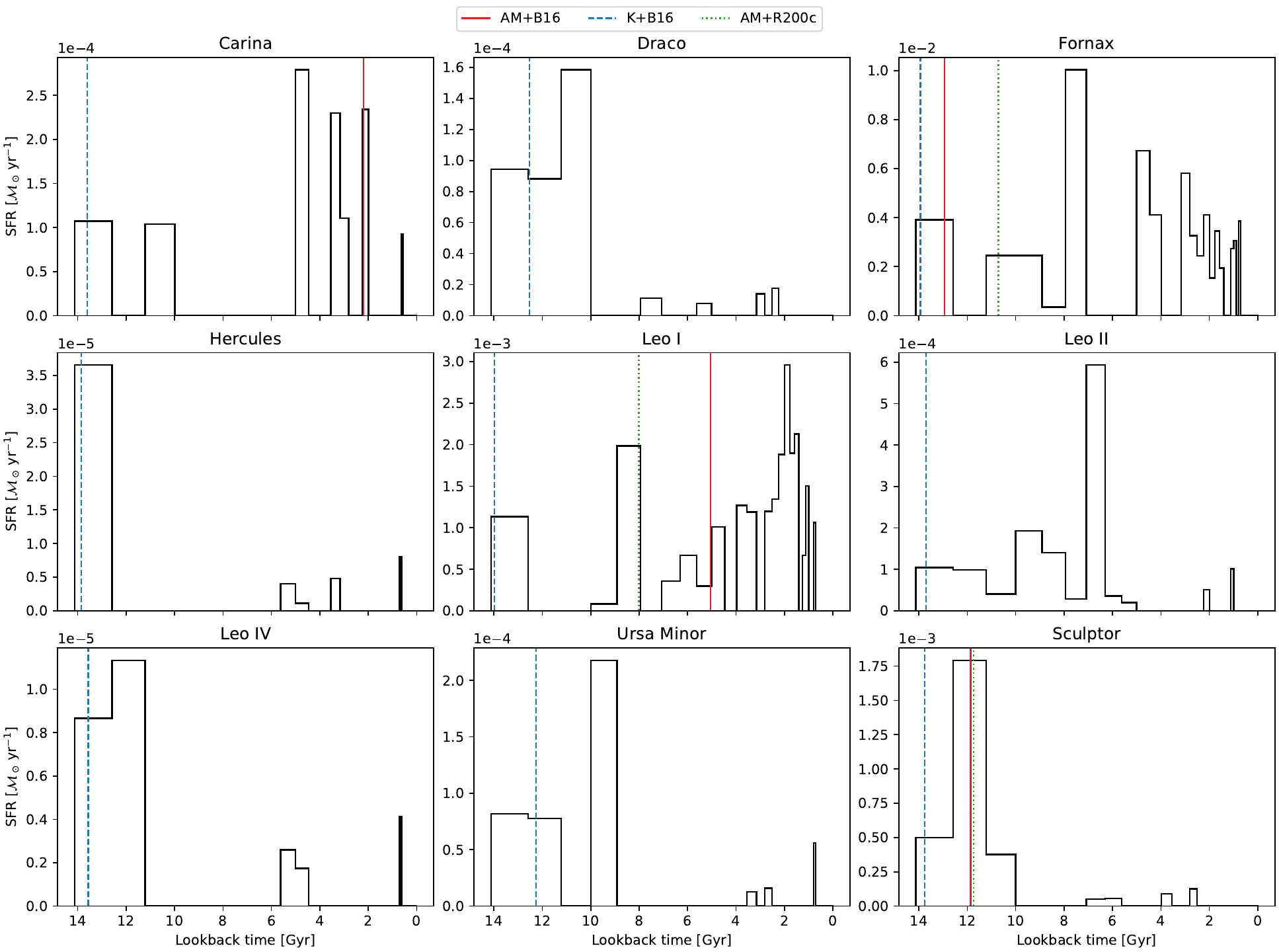}
  \caption{Galaxy-wide differential star formation histories and inferred
  lookback energy-equivalence times $\tenergylb$. Solid, dashed, and dotted
  vertical lines denote AM+B16, K+B16, and AM+$R_{200{\rm c}}$,
  respectively. A vertical line is plotted only when
  $0<f_{\rm crit}^{(p)}\leq1$; no arbitrary time is assigned when the
  threshold is not reached.}
  \label{fig:sfh_energy_equivalence}
\end{figure*}

\begin{table*}
  \centering
  \caption{Normalization and energy-threshold results for the final
  primary sample. A dash denotes that the nominal threshold is not
  reached.}
  \label{tab:energy_threshold_results}
  \begin{tabular}{lcccccccc}
    \hline
    Name & $\mass_{\star,{\rm formed}}^{\rm tot}$ & $\tninetylb$
    & \multicolumn{2}{c}{AM+B16}
    & \multicolumn{2}{c}{K+B16}
    & \multicolumn{2}{c}{AM+$R_{200{\rm c}}$} \\
    & [$\msun$] & [Gyr]
    & $f_{\rm crit}$ & $\tenergylb$ [Gyr]
    & $f_{\rm crit}$ & $\tenergylb$ [Gyr]
    & $f_{\rm crit}$ & $\tenergylb$ [Gyr] \\
    \hline
    Carina & $6.3333\times10^{5}$ & 2.814 & 0.9125 & 2.201 & 0.0875 & 13.604 & 1.4091 & -- \\
    Draco & $4.8333\times10^{5}$ & 10.149 & 1.0972 & -- & 0.3090 & 12.536 & 1.6050 & -- \\
    Fornax & $3.3333\times10^{7}$ & 2.235 & 0.1393 & 12.933 & 0.0227 & 13.928 & 0.2172 & 10.712 \\
    Hercules & $6.1667\times10^{4}$ & 12.602 & 1.1776 & -- & 0.1566 & 13.857 & 4.3478 & -- \\
    Leo I & $9.1667\times10^{6}$ & 1.678 & 0.5086 & 5.051 & 0.0184 & 13.973 & 0.3948 & 8.009 \\
    Leo II & $1.2333\times10^{6}$ & 6.410 & 1.1168 & -- & 0.0355 & 13.702 & 1.0231 & -- \\
    Leo IV & $3.1667\times10^{4}$ & 11.244 & 2.2018 & -- & 0.1516 & 13.567 & 6.0104 & -- \\
    Ursa Minor & $4.8333\times10^{5}$ & 9.064 & 1.3397 & -- & 0.3120 & 12.262 & 1.6050 & -- \\
    Sculptor & $3.8333\times10^{6}$ & 10.606 & 0.5421 & 11.854 & 0.0477 & 13.755 & 0.5957 & 11.739 \\
    \hline
  \end{tabular}
\end{table*}

\subsubsection{Dependence on the Halo Prescription}
\label{sec:result_halo_dependence}

The three threshold estimates combine empirical halo-mass and
halo-radius prescriptions; they are not statistically independent
measurements. AM+B16 and AM+$R_{200{\rm c}}$ share the same
abundance-matching mass, whereas AM+B16 and K+B16 share the
Brook--Shankar size relation.

The final sample divides as follows. (i) Fornax, Leo~I, and Sculptor
reach the nominal threshold under all three prescriptions, with
$f_{\rm crit}^{(p)}$ between 0.018 and 0.60; even for these galaxies,
the crossing epochs differ by up to $\sim$8 Gyr among prescriptions
(Table~\ref{tab:energy_threshold_results}). (ii) The remaining six
galaxies reach the threshold only under some prescriptions. Carina
reaches it under AM+B16 ($f_{\rm crit}=0.91$) and K+B16
($f_{\rm crit}=0.088$) but not under AM+$R_{200{\rm c}}$
($f_{\rm crit}=1.41$). Draco, Hercules, Leo~II, Leo~IV, and Ursa Minor
reach it only under K+B16, with $f_{\rm crit}$ between 0.036 and 0.31,
while their abundance-matching values exceed unity
($1.02\leq f_{\rm crit}\leq6.0$). (iii) No galaxy fails under every
prescription, mainly because K+B16 gives systematically lower halo
masses.

Agreement among prescriptions shows reduced sensitivity to these
mappings; it does not prove that a wind occurred. Failure to reach the
threshold also does not prove that a wind was absent.

\subsubsection{Relation to the SFH Completion Time}
\label{sec:result_sfh_comparison}

Under K+B16, every galaxy crosses the threshold at lookback times of
12.3--14.0 Gyr, within the earliest one or two SFH bins, and always
before $\tninetylb$. For Hercules, whose star formation was essentially
complete in the oldest bin, the K+B16 crossing and $\tninetylb$ fall in
the same SFH bin. Under the abundance-matching prescriptions, existing
crossings also precede $\tninetylb$ for Fornax, Leo~I, and Sculptor,
with separations of 1.1--10.7 Gyr in lookback time. The only exception
is Carina under AM+B16: the threshold is reached at 2.2 Gyr lookback
time, later than $\tninetylb=2.8$ Gyr, so Carina had already formed
90\% of its stars before the nominal energy condition was met. These
comparisons are descriptive: because $\tenergylb$ and $\tninetylb$ come
from the same cumulative SFH, their ordering or closeness in time is
not independent evidence of feedback-driven quenching.

Continued or renewed star formation after the nominal threshold shows
that $\esn=\ebind$ cannot be identified with permanent quenching.
Possible explanations include incomplete gas removal, radiative losses,
outflow followed by reaccretion, spatially uneven feedback, or
continued gas accretion.

\subsection{Auxiliary Luminosity--Metallicity Relation}
\label{sec:result_lzr}

\begin{figure}
  \centering
  \includegraphics[width=0.7\columnwidth]{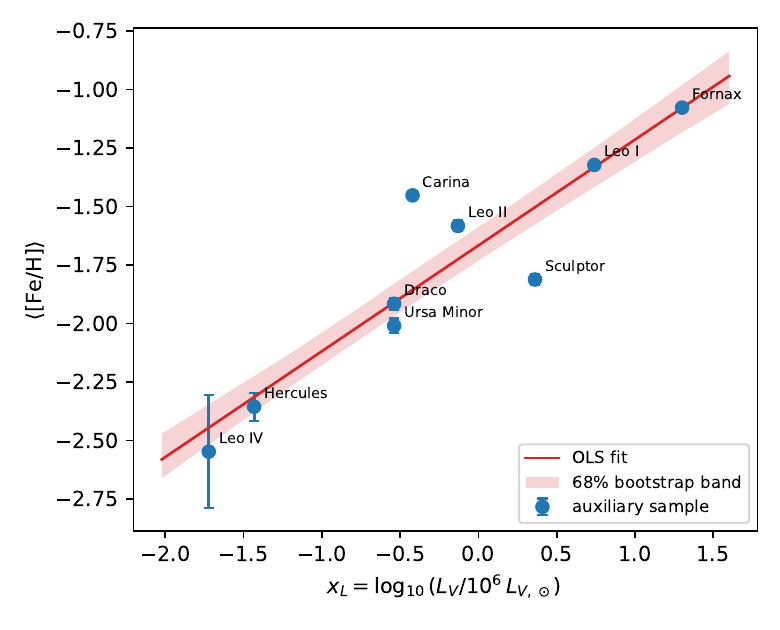}
  \caption{Check on the primary sample: mean stellar [Fe/H] versus
  integrated $V$-band luminosity for the auxiliary sample
  (Section~\ref{sec:data_metallicity}), compared with
  \citet{2013ApJ...779..102K}.}
  \label{fig:lzr}
\end{figure}

We use this relation to check whether our nine galaxies are typical
Local Group dSphs in luminosity and metallicity, rather than an
unusual subsample chosen only because SFH data are available
(Section~\ref{sec:introduction}). Figure~\ref{fig:lzr} shows that they
follow the expected sequence, so the energy analysis is applied to
normal dwarf satellites. This relation is not used in the energy
calculation and is not used to infer outflows.

We write the fitted relation as
\begin{align}
  \left\langle[\mathrm{Fe/H}]\right\rangle
  =
  a_L+b_L\log_{10}\left(\frac{L_V}{10^6L_{V,\odot}}\right) .
  \label{eq:lzr_result_model}
\end{align}
The refitted coefficients are
\begin{align}
  a_L
  &=
  -1.667^{+0.078}_{-0.065},
  \label{eq:lzr_intercept_result}\\
  b_L
  &=
  0.452^{+0.038}_{-0.053}.
  \label{eq:lzr_slope_result}
\end{align}
The corresponding relation of \citet{2013ApJ...779..102K} is
\begin{align}
  \left\langle[\mathrm{Fe/H}]\right\rangle
  =
  (-1.68\pm0.03)
  +(0.29\pm0.02)
  \log_{10}\left(\frac{L_V}{10^6L_{V,\odot}}\right) .
  \label{eq:kirby_lzr}
\end{align}
The refitted intercept agrees with \citet{2013ApJ...779..102K} within
the bootstrap uncertainty, but the slope is steeper than their
$0.29\pm0.02$ and is not statistically consistent with it: the
difference of $\simeq0.16$ exceeds three times the combined uncertainty.
This steepening is plausibly driven by the heterogeneous,
spectroscopy-limited SAGA selection, in which the faintest systems are
represented by small numbers of preferentially metal-poor stars, rather
than by a physical difference from the \citet{2013ApJ...779..102K}
sample. We find no flattening of the relation at low luminosity in the
audited sample and therefore do not discuss a low-luminosity metallicity
floor. The heterogeneous SAGA selection is not fully captured by the
formal regression uncertainty, so we do not use this relation to
establish a metallicity floor or to identify galactic outflows.

\subsection{Physical Interpretation and Limitations}
\label{sec:result_interpretation}

The energy comparison tests whether the observed stellar population
had produced enough nominal supernova energy to match the adopted gas
binding energy by a given epoch. The binding energy depends strongly on
the halo mass, halo radius, density profile, retained gas fraction, and
structure factor. The halo prescriptions are present-day-equivalent
proxies and do not reconstruct halo growth, pre-infall mass, tidal
stripping, or mergers.

The parameter dependence can be summarized analytically. If the halo mass
and radius are treated as independent inputs, then
\begin{align}
  \ebind
  &\propto
  \alpha_{\rm bind}f_{\rm ret}f_{\rm b}
  \mass_{\rm h}^{2}R_{\rm h}^{-1},
  \label{eq:binding_energy_general_sensitivity}\\
  \mass_{\star,{\rm crit}}
  &\propto
  \alpha_{\rm bind}f_{\rm ret}f_{\rm b}
  \mass_{\rm h}^{2}R_{\rm h}^{-1}
  \eta_{\rm SN}^{-1}\epssn^{-1}C_{\rm SN}^{-1},
  \label{eq:critical_mass_general_sensitivity}\\
  f_{\rm crit}
  &\propto
  \alpha_{\rm bind}f_{\rm ret}f_{\rm b}
  \mass_{\rm h}^{2}R_{\rm h}^{-1}
  \eta_{\rm SN}^{-1}\epssn^{-1}C_{\rm SN}^{-1}
  (1-r)\mass_{\star,V}^{-1}.
  \label{eq:critical_fraction_general_sensitivity}
\end{align}
The halo mass and radius should not, however, be varied independently when a
radius prescription ties them together. The Brook--Shankar and
spherical-overdensity mappings imply
\begin{align}
  \ebind^{\rm B16}
  &\propto
  \mass_{\rm h}^{2.49}\reff^{-1},
  \label{eq:binding_energy_b16_sensitivity}\\
  \ebind^{200{\rm c}}
  &\propto
  \mass_{\rm h}^{5/3}H_0^{2/3}.
  \label{eq:binding_energy_r200c_sensitivity}
\end{align}
For the K+B16 closure proxy, eliminating the derived halo mass and radius in
favor of the measured inputs gives
\begin{align}
  \ebind^{\rm K+B16}
  &\propto
  \sigmalos^{3.342282}
  \reff^{0.671141}
  \alpha_{\rm vir}^{-1.671141}.
  \label{eq:binding_energy_kb16_observable_sensitivity}
\end{align}
At fixed halo prescription, adopting
$C_{\rm SN}=7.5354\times10^{-3}\,\msun^{-1}$ instead of the fiducial
value multiplies both $\mass_{\star,{\rm crit}}$ and $f_{\rm crit}$ by
1.47729. Likewise, $\eta_{\rm SN}=0.3$ and 0.1 multiply them by 3.3333 and
10, respectively. Relative to $r=0.40$, the choices $r=0.35$ and 0.45
multiply $f_{\rm crit}$ by 1.08333 and 0.916667, respectively. A smaller
retained gas fraction reduces the binding-energy requirement linearly.

Among the adopted sensitivity cases, coupling efficiency dominates.
Changing the return fraction ($r=0.35$--$0.45$) or varying $\sigmalos$
by $\pm1\sigma$ changes no reached/not-reached outcome except one
marginal case (Leo~II under AM+$R_{200{\rm c}}$ at $r=0.45$, with
$f_{\rm crit}=0.94$), and shifts the crossing bin in at most a few
galaxy--prescription combinations. Restricting $C_{\rm SN}$ to
progenitors of 8--18$\,\msun$ flips only the marginal Carina AM+B16
case ($f_{\rm crit}=0.91\rightarrow1.35$). By contrast,
$\eta_{\rm SN}=0.3$ removes the marginal K+B16 crossings of Draco and
Ursa Minor ($f_{\rm crit}\rightarrow1.03$--$1.04$) and all
abundance-matching crossings except those of Fornax, and
$\eta_{\rm SN}=0.1$ leaves only five galaxies (Carina, Fornax, Leo~I,
Leo~II, and Sculptor) reaching the threshold, each only under K+B16.
The reached/not-reached classification is therefore robust to the
stellar-population and kinematic inputs at SFH-bin resolution, but it
depends on order-unity energy coupling.

The fiducial choice $\eta_{\rm SN}=1$ maximizes the energy available to
the ISM for a fixed number of supernovae. Lower coupling efficiencies
increase $f_{\rm crit}^{(p)}$, shift $\tenergylb^{(p)}$ toward smaller
lookback times, or remove a crossing. The inferred energy-equivalence
time therefore depends on both feedback and gas-retention assumptions.

The CMD-based SFHs have finite time resolution, especially at old ages,
and may be affected by HST field coverage and radial population
gradients. Because $\tenergylb$ and $\tninetylb$ come from the same
cumulative SFH, their closeness in time is not an independent causal
test.

Within these limitations, a threshold reached before or near
$\tninetylb$ is consistent with stellar feedback contributing to gas
redistribution or loss. It does not prove that feedback caused
quenching, and a noncrossing does not exclude outflows under a
different gas, halo, IMF, or coupling model. Environmental processes
remain viable alternatives or complements, but we do not quantify their
separate contributions.

\section{Conclusions}\label{sec:conclusion}

We applied an order-of-magnitude energy-budget test to nine Milky Way
dwarf satellites with published SFHs and galaxy-level kinematic and
structural data. The sample selection is based on data availability
rather than statistical representativeness. When the nominal threshold
is reached, we define $\tenergy$ as the first elapsed time at which
the cumulative effective supernova energy equals the adopted gas
binding energy, with $\tenergylb$ as its lookback-time counterpart.
Our main finding is that star formation often continues long after
this energy-equivalence time, separating threshold crossing from the
permanent cessation of star formation.

Under the fiducial parameters, Fornax, Leo~I, and Sculptor reach the
nominal threshold under all three halo prescriptions; Carina reaches
it under AM+B16 and K+B16; and Draco, Hercules, Leo~II, Leo~IV, and
Ursa Minor reach it only under K+B16. Thus, all nine galaxies reach
the threshold under at least one prescription. The distinction between
energetic sufficiency and quenching is evident in the timing comparison:
under K+B16, every galaxy crosses the threshold at lookback times of
12.3--14.0 Gyr, always before $\tninetylb$, yet most systems continued
forming stars for many Gyr afterward. Fornax, Leo~I, and Carina, for
example, formed most of their stars after the crossing. Under the
abundance-matching prescriptions, crossings also usually precede
$\tninetylb$, except for Carina under AM+B16. The reconstructed SFHs
satisfy
$\sum_i\sfr_i\Delta\tau_{i,{\rm yr}}=
\mass_{\star,{\rm formed}}^{\rm tot}$ to numerical precision, with
$\max|\delta_{\rm norm}|=1.2\times10^{-16}$.

The timing of gas loss and the cessation of star formation therefore
depend on additional physical processes. Potential contributors
include cosmic reionization
\citep{2000ApJ...539..517B,2001PhR...349..125B},
ram-pressure stripping and tidal effects after infall
\citep{1972ApJ...176....1G,2020ApJ...905..109M},
and incomplete or delayed coupling of supernova energy to the gas.
Their relative contributions remain undetermined in this analysis.

Figure~\ref{fig:lzr} confirms that the nine galaxies follow the usual
luminosity--metallicity relation of Local Group dSphs
\citep{2013ApJ...779..102K}, so the energy sample is not unusual in
chemistry or luminosity.

The quantitative threshold estimates are affected by uncertainties in
halo mass, halo radius, retained gas fraction, structure factor,
stellar-mass normalization, core-collapse fraction, and
supernova--gas coupling efficiency. The adopted halo estimates describe
present-day-equivalent scales rather than pre-infall histories, and
the temporal interpretation is limited by the finite resolution of
the published SFHs. Taken together, the results distinguish nominal
energetic sufficiency from the physical processes that determine
when star formation ceases.

\section*{Acknowledgments}

We thank Masashi Chiba for valuable comments on the mass-estimation
calculations and the data used. We also thank Kai T.\ Kono and Shun-ya S.\ Uchida for invaluable
help with the calculations and for fruitful discussions. This work was
supported by the Japan Society for the Promotion of Science (JSPS)
Grants-in-Aid for Scientific Research (19H05076, 21H01128, 21J23611,
JP23KJ1089, and 24H00247) and by JST CREST Grant Number JPMJCR24Q1.
This work was also supported in part by the Sumitomo Foundation Fiscal
2018 Grant for Basic Science Research Projects (180923) and by the
Collaboration Funding of the Institute of Statistical Mathematics,
``Machine-Learning-Based Cosmogony: From Structure Formation to Galaxy
Evolution.'' A large language model was used to assist with language
editing, algebraic and numerical consistency checks, and development of
analysis and test code. The authors remain responsible for verifying all
scientific assumptions, calculations, code, interpretations, and final
wording.


\bibliographystyle{mnras}
\bibliography{dwarf_galaxy_wind}

\end{document}